\documentclass[cameraready]{vgtc}

\graphicspath{{figures/}{pictures/}{images/}{./}} 

\usepackage{times}                     

\usepackage{tabu}                      
\usepackage{booktabs}                  
\usepackage{lipsum}                    
\usepackage{mwe}                       

\usepackage{ulem}
\usepackage{multirow}
\usepackage{amsmath}
\usepackage{graphicx}
\usepackage{makecell}
\usepackage{array} 
\usepackage{ulem}
\renewcommand{\arraystretch}{1.2}
\onlineid{0}

\vgtccategory{Research}

\vgtcinsertpkg

\title{Zeitgebers-Based User Time Perception Analysis and Data-Driven Modeling via Transformer in VR}

\author{Yi Li\thanks{e-mail: elieli0925@qq.com} %
\and Zengyu Liu\thanks{e-mail: liuzengyu1209@gmail.com} %
\and Xiandi Zhu\thanks{e-mail:antimony\_51@163.com}
\and Ning Xie\thanks{Corresponding author, e-mail: seanxiening@gmail.com}}
\affiliation{\scriptsize Center for Future Media, the School of Computer Science and Engineering, UESTC, Chengdu, China}

\teaser{
  \centering
  \includegraphics[width=\linewidth]{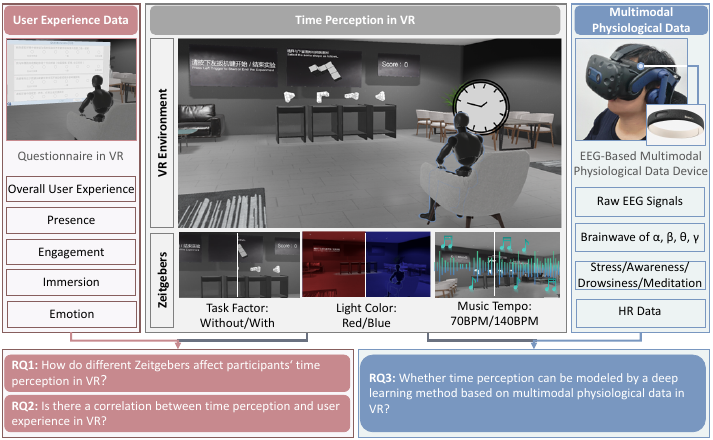}
  \caption{ Overview of the VR Time Perception Study Framework. The center of the image illustrates the use of different zeitgebers (task factor, light color, music tempo) to manipulate time perception within the VR environment. The left section shows the methods used to measure user experience, focusing on dimensions such as presence, engagement, immersion, and emotion. The right section highlights the experimental equipment and physiological data collected, including EEG signals, brainwave data, human state signals, and heart rate (HR). The three research questions addressed in this study are summarized at the bottom.}
  \label{fig:teaser}
}

\abstract{
    Virtual Reality (VR) creates a highly realistic and controllable simulation environment that can easily manipulate users' sense of space and time. While the sensation of ``losing track of time" is often associated with enjoyable experiences, the link between time perception and user experience in VR and the underlying mechanisms remains largely unexplored. 
    This study investigated how different zeitgebers—light color, music tempo, and task factor—influence time perception. And We introduced the Relative Subjective Time Change (RSTC) method to explore the relationship between time perception and user experience. Additionally, we applied a data-driven approach called the Time Perception Modeling Network (TPM-Net), which integrates Convolutional Neural Network (CNN) and Transformer architectures to model time perception based on multimodal physiological and zeitgebers data.    
    With 56 participants in a between-subject experiment, our results show that task factors significantly influence time perception, with red light and slow-tempo music further contributing to time underestimation. The RSTC method effectively reveals that relatively underestimating time in VR is strongly associated with an improved overall user experience, presence, and engagement. Furthermore, the TPM-Net based on multimodal data has shown great potential for modeling time perception in VR, enabling further inference of relative changes in users' time perception and corresponding changes in user experience. 
    This study provides valuable insights into the relationship between time perception and user experience in VR, with the promising time perception model via the TPM-Net data-driven method. This study sheds light on time perception applications in VR-based therapy and specialized sector training. 
} 


\begin{document}

\firstsection{Introduction}
\maketitle

Time is a crucial dimension that shapes our perception and understanding of the world~\cite{burnham2007kant}. In VR, this fundamental concept of time becomes even more complex, as users' time perception can be significantly influenced by immersive zeitgebers. At the same time, the phrase ``time flies" is often associated with positive experiences, a phenomenon linked to the flow state in positive psychology~\cite{seligman2000positive}. In gaming, losing track of time is frequently linked to enjoyable experiences~\cite{sackett2010you, rutrecht2021time, cassidy2010effects}.
Most existing studies focus on analyzing how virtual zeitgebers—such as sun movement, virtual avatar, or task complexity factors—affect time perception in VR~\cite{sabat2022cognitive, mizoguchi2023effect}. No research has further explored the relationship between time perception and user experience in VR. Moreover, these studies typically use physical time as the standard for comparison with subjective time estimates~\cite{walker2022impact}. Most importantly, few individuals can accurately estimate physical time~\cite{walker2022impact}, and we cannot directly link accurate physical time estimation to standardized user experience. Therefore, this study employs visual (light color)~\cite{ cha2020effects}, auditory (music tempo)~\cite{ross2022time, cassidy2010effects}, and task-related~\cite{livesey2007time, read2021engagement} zeitgebers as conditions for influencing users' time perception and proposes the Relative Subjective Time Change (RSTC) method for an in-depth discussion of the relationship between users' relative changes of time perception and user experience.

Furthermore, to better understand time perception mechanisms in VR, we modeled time perception based on multimodal data from users' physiological signals and virtual zeitgebers. Previous studies have shown that physiological states (e.g., EEG, HR, human state, etc.) can reflect users' time-related cognitive behaviors~\cite{silva2022bromazepam, damsma2021temporal,ghaderi2018time}. However, few studies have applied multimodal data-driven methods to model time perception in VR. To address this gap, we propose the Time Perception Modeling Network (TPM-Net), which excels at capturing both complex local temporal features and global dependencies in physiological data~\cite{yang2022mobile}. TPM-Net enables us to analyze users' time perception in VR, infer changes in time perception across different conditions, and potentially speculate changes in user experience.

Time perception is not only closely tied to user experience but also related to mental health conditions (e.g., mood disorder, ADHD, etc.) ~\cite{droit2013time,nejati2020time}. A deeper exploration of time perception can guide VR design and has potential applications in VR-based therapy and specialized training.

In summary, this study explores the effects of various zeitgebers in VR on time perception and analyzes the relationship between time perception and user experience. We propose the TPM-Net framework for time perception modeling in VR by integrating multimodal data and deep learning techniques. The contributions of this paper are summarized as follows:

(1) We categorized zeitgebers into visual, auditory, and task factors, focusing on common design elements such as color, music tempo, and task factor. Using a qualitative research approach, we explored how these elements influence time perception, enriching the understanding of time perception within VR environments.
    
(2) We conducted an in-depth analysis of the relationship between time perception and user experience, introducing the novel Relative Subjective Time Change (RSTC) method. By leveraging relative changes in both time perception and user experience, this method more effectively validates experimental conclusions, offering more substantial empirical support.

(3) We introduced a multimodal physiological data-driven and the TPM-Net to model user time perception in VR. This new data-driven method allows for accurate time perception modeling and offers a novel approach for predicting relative changes in user time perception and user experience.

\section{Related Work}
This section begins with an introduction to the concept of time perception and the psychological principles influencing it. We then review research on time perception in VR, highlighting the potential relationship between time perception and user experience. Finally, we examine time perception measurement paradigms and related data-driven approaches to lay the groundwork for subsequent experimental design and deep learning methods.  
\subsection{Time Perception}
The concept of time can be divided into biological and psychological time~\cite{grondin2001physical}. Biological time evolves as an adaptation of living creatures to the earth's day-night cycles~\cite{koukkari2007introducing}. Psychological time includes temporal-duration, temporal-sequence, and temporal-perspective perception~\cite{ivry2008dedicated}. Temporal-duration perception refers to the subjective estimation of time intervals or event durations. In this work, ``time perception" refers to estimating subjective time duration. The term zeitgeber originates from German and means time-giver, influencing human time perception~\cite{sharma2005zeitgebers, duffy2005entrainment}. External zeitgebers are environmental cues that affect the body's biological rhythms, while internal clocks are innate biological rhythms within humans or other organisms~\cite{menna2012external}.

How do zeitgebers affect time perception? Several popular theories are trying to explain the mechanisms of time perception. One theory, the pacemaker-accumulator model, explains how emotions and arousal shape subjective time~\cite{ wearden1995feeling, droit2007emotions}. 
In this model, an internal clock in our body consists of a pacemaker, a switch, and an accumulator. The pacemaker emits pulses, the switch controls pulse entry, and the accumulator collects pulses. More pulses mean longer perceived time~\cite{smith2011effects, cui2023role}. Colors influence arousal and thus time perception, with studies showing that long-wavelength colors like red can increase arousal, while short-wavelength colors like blue have a calming effect. In turn, being in different light colors can affect time duration~\cite{yang2018subjective,cha2020effects}.

The Attentional-Gate Model (AGM)~\cite{block1996models,krug2024transient} has become widely accepted with the development of cognitive science. 
In addition to the pacemaker-accumulator model, AGM adds attention and working memory to the original pacemaker-accumulator framework. ~\cite{van2011psychological}.
When attention is on time, more pulses are generated and collected relatively faster, expanding time perception\cite{sackett2010you, taylor1994waiting}.
This implies that task factors consuming attentional resources also influence time perception.~\cite{barrouillet2007time, read2021engagement}.

Another theory often used to explain time perception in music is entrainment~\cite{ross2022time}. Entrainment refers to the temporal coupling or synchronization of two independent oscillatory systems through phase alignment. Neural entrainment causes the brain's neural oscillations to synchronize with musical rhythms, helping the brain predict and prepare for the rhythm before it occurs, thus influencing time perception~\cite{ross2022time,cassidy2010effects}.

In general, elements like color, task factor, and music are commonly used in VR, but their effects on time perception are underexplored. Understanding these influences can improve VR design and our grasp of time perception in immersive environments.

\subsection{User Experiment and Time Perception in VR}
There seems to be a link between user experience and time perception, supported by several studies~\cite{seligman2000positive, csikszentmihalyi2005flow, sackett2010you, rutrecht2021time}. 
 User experience refers to the perceptions and responses from using or anticipating the use of a system, product, or service~\cite{dis20099241}. 
 In daily life, we often say ``time flies" during positive experiences, a phenomenon linked to the flow state in positive psychology~\cite{seligman2000positive}. Flow is when individuals are fully immersed, experiencing joy, focus, and satisfaction~\cite{csikszentmihalyi2005flow}. A distorted sense of time is a crucial feature of flow~\cite{csikszentmihalyi2005flow}.
Overestimating time means perceiving it as longer than it is, while underestimating reflects a shorter perception~\cite{pande2010overestimation}. Time underestimation often correlates with positive experiences. In gaming, losing track of time is linked to enjoyable experiences~\cite{sackett2010you, rutrecht2021time, cassidy2010effects}.
Sackett et al. showed that when people felt that time went by faster than expected, they found tasks more enjoyable, noises less annoying, and songs more catchy~\cite{sackett2010you}. 
Rutrecht et al.\cite{rutrecht2021time} showed that strong flow in games led to less focus on time, with time passing faster, especially in VR.
Cassidy et al.\cite{cassidy2010effects} found that participants overestimated time with self-selected music in a driving game but underestimated time with high-arousal music, which increased cognitive load and reduced enjoyment. 
In summary, it is reasonable to hypothesize that there is a potential relationship between time perception and user experience.

However, many VR time perception studies focus on static factors, with little attention to their effects on user experience.  
Some have explored zeitgebers such as virtual embodiment~\cite{unruh2023body, mizoguchi2023effect}, music tempo~\cite{ picard2023rhythmic, rogers2020potential}, moving visual elements~\cite{sabat2022cognitive, schatzschneider2016turned} and virtual pendulums~\cite{landeck2023clocks}, examining their effects on time perception in VR.
Other research compares zeitgebers effects on time perception in real and virtual settings~\cite{van2019elapsed, liao2020data, landeck2023clocks}. 
Only a few studies explore the relationship between user experience and temporal awareness in the real world~\cite{van2019elapsed, liao2020data, landeck2023clocks}. Landeck et al.\cite{landeck2023clocks} found a significant negative correlation between time perception and boredom. Picard et al.\cite{picard2023rhythmic} showed participants' performance aligned with their time perception, with additional stimuli making time seem faster. Schatzschneider et al.\cite{schatzschneider2016turned} found that presence influenced time estimation, though effects were insignificant. Rogers et al.\cite{rogers2020potential} explored the impact of music on time perception and immersion in a VR game, finding that while fast time perception was often linked to immersion, the two were not always correlated, suggesting a need to reevaluate this relationship.
These studies primarily examine individual aspects of user experience, like immersion, presence, and emotion. However, user experience in VR is multidimensional, including flow, engagement, and usability~\cite{tcha2016questionnaire, wienrich2018assessing}. This paper analyzes the relationship between time perception and overall user experience in VR, including its components.

\subsection{Measurements of Time Perception}
Methods for measuring time perception can be categorized into subjective and objective approaches. In this paper,  the subjective approach involves estimating time through psychological time estimation paradigms~\cite{zakay1993relative,zhen2006effects}, and the objective approach entails directly assessing time perception using multimodal physiological data~\cite{johari2023temporal,fontes2016time, silva2022bromazepam,damsma2021temporal} such as electroencephalogram (EEG) and heart rate (HR).

There are two primary time estimation paradigms: the prospective and the retrospective. In the prospective paradigm, users are informed in advance that they need to estimate the duration, making it reliant on attentional resources~\cite{zakay1993relative}. In contrast, the retrospective paradigm asks users to estimate time only after the task is complete, relying on working memory. Studies show that the prospective paradigm is generally more accurate and demands greater attentional resources than the retrospective one~\cite{zhen2006effects}. Given these paradigms, we adopt the prospective approach in this paper. However, in the prospective paradigm, there are two main methods: participant self-report and event-triggered responses. In self-report, participants inform the experimenter or input data via an interface~\cite{read2023influence, liao2020data, van2019elapsed}. In event-triggered responses, participants press a button when they believe the required duration is reached~\cite{unruh2023body, mizoguchi2023effect}. Due to potential inaccuracies in reporting times below one second, this study adopts the event-triggered method.

For the objective method, studies have shown that EEG and other physiological data can effectively represent and reflect time perception.  Fontes et al.\cite{fontes2016time} explored the relationship between the brain and time perception, concluding that regions like the frontal cortex, parietal, basal ganglia, cerebellum, and hippocampus play key roles in this process. Silva et al.\cite{silva2022bromazepam} focused on temporal estimation and the impact of EEG alpha asymmetry in the frontal and motor cortex. Other studies highlight the importance of beta and theta frequency subbands in time perception~\cite{ damsma2021temporal}. Ghaderi et al.\cite{ghaderi2018time} found that participants who overestimated time showed lower beta band activity, which also correlated with increased cognitive complexity. Additionally, HR has been shown to be a crucial indicator of time-perception conditions\cite{ogden2022psychophysiological}. Moreover, time perception is a complex, multi-layered cognitive experience, with human states—such as physiological and psychological conditions—providing key insights into how time is perceived~\cite{tamm2014compression}.
In this paper, we leverage EEG, brainwave data, and human states derived from EEG and HR as multimodal physiological inputs, combined with the TPM-Net, to model participants' time perception.

\section{MATERIAL AND METHODS}
This chapter introduces the experimental approach. Section 3.1 provides an overview of the research questions and methodology, Section 3.2 details the design of experiments, and Section 3.3 describes the architecture of the data-driven modeling network. The notations used in this paper are summarized in Table~\ref{table:terms description}. 
\begin{table}[htbp]
  \centering
  \caption{ Parameters Summary.}
\begin{tabular}{lp{5.7cm}}
    \toprule
    \textbf{TERMS} & \multicolumn{1}{l}{\textbf{DESCRIPTIONS}} \\
    \midrule
    \multicolumn{1}{p{3.66em}}{$t$} & subjective time estimation \\
    \multicolumn{1}{p{3.66em}}{$\hat{t}$} & mean of subjective time estimates for repeated experiments \\
    \multicolumn{1}{p{3.66em}}{$\Delta \hat{t}$} & mean of subjective time estimation gap compared to corresponding baseline scenario \\
    \midrule
    \multirow{2}[1]{*}{$cl$} & cl = 1: with cognitive task \\
          & cl = 0: without cognitive task \\
    \multirow{2}[0]{*}{$c$} & c = 1: blue light scenario \\
          & c = 2: red light scenario \\
    \multirow{2}[1]{*}{$m$} & m = 1: 140 BPM music \\
          & m = 2: 70 BPM music \\
    \bottomrule
    \end{tabular}%
    
  \label{table:terms description}%
\end{table}%

\subsection{Research Question}
To deeply explore the mechanisms by which different zeitgebers affect time perception in VR environments and analyze the relationship between time perception and user experience, we formulated the following three key research questions:

\textbf{RQ1:} How do different zeitgebers affect participants' time perception in VR?

\textbf{RQ2:} Is there a correlation between time perception and user experience in VR?

\textbf{RQ3:} Can time perception be effectively modeled by a deep learning approach based on multimodal physiological data in VR?

To address RQ1, we categorized zeitgebers into visual, auditory, and task factors, selecting light color~\cite{yang2018subjective,cha2020effects}, music tempo~\cite{ross2022time, cassidy2010effects}, and task factor~\cite{read2021engagement,block1997prospective} as the key zeitgebers influencing time perception. These zeitgebers were chosen not only because of their practical relevance in VR applications but also due to their well-established roles in shaping time perception. For RQ2 and RQ3, we focused on zeitgebers that primarily influence time perception, as RQ3 aims to model time perception using TPM-Net based on multimodal data rather than user experience. And this also allows us to more accurately capture the intricate relationship between time perception and user experience.

In our experimental design, we employed a prospective paradigm, informing participants in advance that they would need to estimate the 
time duration (60 seconds) and allowing them to trigger the end of the timing during the experiment actively. We chose this method because it accurately reflects participants' time perception, thus offering more reliable data for our study~\cite{unruh2023body,zakay1993relative}.

To address RQ2, participants completed a user experience questionnaire in VR after each scenario, assessing presence, engagement, immersion, emotions, and overall experience~\cite{tcha2016proposition}. Completing the questionnaire within VR helped avoid disruptions from returning to the real world~\cite{regal2019questionnaires}. We established a baseline for each experiment group and analyzed correlations between relative changes in time perception and user experience with the RSTC. This approach not only addresses the challenge of linking physical time accuracy with user experience but also mitigates the effects of factors such as age and gender~\cite{siu2014time, gagnon2021age}, strengthening the persuasiveness of our findings.

Before the formal experiment, we conducted a full-factorial pilot study with 12 participants. In the pre-experiment, participants frequently engaged in trials where the differences in user experience were minimal, making it challenging to discern variations in user experience questionnaire scores. Additionally, task factors had a dominant impact on time perception, overshadowing the effects of visual and auditory factors. As a result, we employed a mixed single-subject repeated measures design across four groups for the formal experiment~\cite{wambaugh2014single}, as shown in Table~\ref{table:Experimental groups and sequence}. Participants were randomly assigned to one of the four groups and performed the corresponding Baseline experiment.
\begin{table}
\centering
\caption{Experimental groups. Participants need to complete the corresponding baseline experiment and one other group experiment. For color parameters, 0 indicates white light, 1 indicates blue light, and 2 indicates red light. For music parameters, $\emptyset$ indicates no music.}
\renewcommand{\arraystretch}{1.8}
\scalebox{0.75}{
\begin{tabular}{ccccccc}
        \toprule
        \textbf{\#}  & \multicolumn{2}{c}{\textbf{Group Type}} & \textbf{\makecell[c]{Color\\Parameter($c$)}} & \textbf{\makecell[c]{Music\\Parameter($m$)}} &\textbf{Terms} & \\
        \cline{1-7}
        0 &\multirow{5}{*}{\rotatebox{90}{With Task  }} & Baseline  & 0 & $\emptyset$ & $G_{baseline}^{cl=1}$   &\\
        \multirow{2}{*}{1}& & \multirow{2}{*}{Color Group} &1 & $\emptyset$ & $G_{c=1}^{cl=1}$  & \multirow{2}{*}{$\Bigg\}$ \makecell[c]{Random \\Order}} \\
        & & &2 & $\emptyset$& $G_{c=2}^{cl=1}$& \\

         \multirow{2}{*}{2} & &\multirow{2}{*}{Music Group} & 0 & 140BPM& $G_{m=1}^{cl=1}$  & \multirow{2}{*}{$\Bigg\}$ \makecell[c]{Random \\Order}}\\
         & & & 0 & 70BPM& $G_{m=2}^{cl=1}$ &  \\
        \hline
         0*&\multirow{5}{*}{\rotatebox{90}{Without Task  }} & Baseline  & 0 & $\emptyset$& $G_{baseline}^{cl=0}$  & \\
        \multirow{2}{*}{3}& & \multirow{2}{*}{Color Group} & 1 & $\emptyset$& $G_{c=1}^{cl=0}$ &\multirow{2}{*}{ $\Bigg\}$ \makecell[c]{Random \\Order}}  \\
        & & &2 & $\emptyset$& $G_{c=2}^{cl=0}$ & \\

         \multirow{2}{*}{4} & &\multirow{2}{*}{Music Group} & 0 & 140BPM & $G_{m=1}^{cl=0}$ &\multirow{2}{*}{ $\Bigg\}$ \makecell[c]{Random \\Order}}  \\
         & & & 0 & 70BPM& $G_{m=2}^{cl=0}$  & \\
         \bottomrule
\label{table:Experimental groups and sequence}
\end{tabular}
}
\end{table}


To address RQ3, we used non-invasive multimodal physiological data collection devices to gather participants' EEG, brainwave data, human state data, and HR data during the experiment. And then we proposed the TPM-Net method to model VR time perception, which we compared with traditional models to evaluate its performance. Recognizing that accurately timing physical durations is challenging for most people~\cite{walker2022impact}, we classified this multimodal data into three categories (underestimated/acceptably accurate/overestimated) and established a custom time range ([51,69)) as the standard for timing acceptably accuracy (more details in 3.3.).

\subsection{Design of Experiments}

\subsubsection{Participants}
A total of 60 participants took part in the study. Four participants had to be excluded because they did not follow the protocol or had technical difficulties. Of the remaining 56 participants, 21 were females, and 35 were males, all of whom were students ranging in age from 19 to 28 years (M = 23.03, SD = 1.89). 54 participants were right-handed, one was left-handed, and no one was ambidextrous. None of them are color blind or hard of hearing. The experiment was supervised by the XXX Ethics Committee, and all experimental data and information were kept strictly confidential. A VR test with a different scenario than the formal experiment was conducted before recruitment to confirm that participants did not have cybersickness. 12 of 56 participants had already participated in at least one VR study.

\subsubsection{Virtual Reality Experimental Scene Design}
We conducted the experiment in a quiet, comfortable, temperature-controlled laboratory. In the virtual environment, participants were placed in a sealed, modern lounge, where they sat on a chair facing three taskboards displaying tasks and procedures. The room's decor featured neutral tones like black, white, and gray~\cite{wallach1963perception} as neutral colors have minimal impact on the user's state~\cite{kurt2014effects}.  In the baseline scenario, white lighting was used to minimize environmental influence. Below the taskboards, four game boxes displayed interactive task items.

The task was a spatial rotation task. The middle taskboard showed the correct answer, and the game boxes displayed four objects at different rotations. Each shape had four possible rotations, with two randomly selected for each answer. Participants used a toy gun in their right hand to choose the two matching objects from four options. Correct answers earned points; incorrect ones deducted them. The toy gun had a 2-second refresh time to prevent random guessing. The left-hand controller started and ended the experimental trial.

Six sets of ceiling lights controlled the scene's color in the virtual environment. The lights could change to white, blue, or red, with parameters detailed in Table \ref{table:HSVColor}. Light intensity was preset and consistent, and the soft shadow mode remained unchanged during color transitions. Unity's Universal Render Pipeline (URP) optimized the lighting, shadows, and post-processing effects for visual comfort~\cite{tam2011stereoscopic}.

The audio environment included three conditions: no music, fast-tempo music (140 BPM), and slow-tempo music (70 BPM). We collaborated with professional musicians to ensure clear tempo distinctions and a light style to minimize emotional impact\footnote{\url{https://drive.google.com/drive/folders/1w_8or0z8PwR3V8h_1Cu_CCWS4TNZ-yXA?usp=sharing}}. The music was played through a virtual voice box in the VR environment, and the Doppler effect was neutralized to maintain an immersive auditory experience.

The VR experiment was developed by Unity3D2021.3.4f1c1, with HTC VIVE Pro for immersive interaction. The multimodal physiological data acquisition device, BrainCo OxyZen\footnote{\url{https://brainco.tech/}}, captured EEG and HR data, calculating brainwave metrics like $\alpha$, $low\beta$, $high\beta$, $\theta$, and $\gamma$, along with human states such as stress, awareness, drowsiness, and meditation~\cite{han2022patent,han2021patent}. The EEG sampling rate was 256Hz. We developed the data acquisition module using PyCharm and Python, employing UDP communication with Unity to control data recording.

\begin{table}[h]
    \centering
  \caption{HSV-based light color and intensity specific parameters}
  \scalebox{0.9}{
  \begin{tabular}{ccccc}
    \toprule
    \textbf{Light Color} &\textbf{Hue} &\textbf{Saturation} &\textbf{Value}&\textbf{Ligh Intensity}\\
    \midrule
     c = 0 & 0.0000   &0.0000      &1.0000& 11\\
     c = 1  & 0.0000   &0.6983     &0.7020 &11\\
    c = 2& 0.6667   &0.7990     &0.8000 &11\\
    \bottomrule
    \label{table:HSVColor}
  \end{tabular}
  }
\end{table}
\subsubsection{Procedure}

\begin{figure}[!h]
     \centering
  \begin{center}
\includegraphics[width= 0.5\columnwidth]{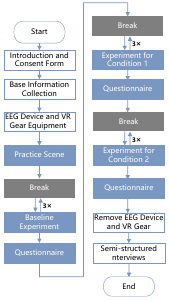}
  \caption{Experiment Process.}
  \label{fig:ExperimentProcess}
    \end{center} 
\end{figure}

\begin{figure}[t]
     \centering
  \begin{center}
\includegraphics[width= 0.8\columnwidth]{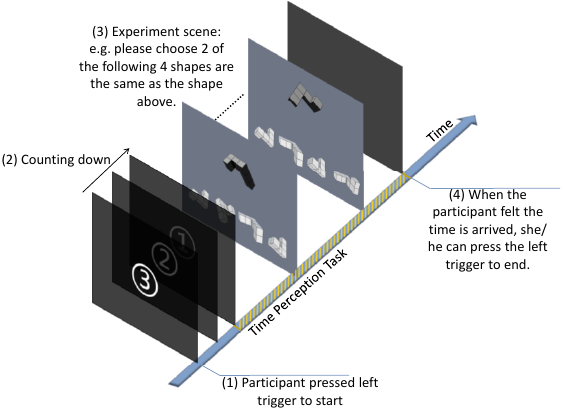}
  \caption{Example of the sequence of one trail. Each trail is divided into four sections. Participants start timed tasks after the countdown has ended.}
  \label{fig:CognitiveLoadExperience}
    \end{center} 
\end{figure}

First, participants were randomly assigned to one of the groups indicated in Table \ref{table:Experimental groups and sequence} (Groups 1 through 4). The experimental process is outlined in Figure \ref{fig:ExperimentProcess}. Before starting the experiment, participants received a full explanation of the study and signed an informed consent form. They were then equipped with a VR headset and EEG devices (as shown in Figure \ref{fig:teaser} right), and the equipment functionality was verified to ensure proper operation. Participants first entered a practice scenario to fully understand and practice the spatial rotation task. This step was skipped for groups without tasks.
Next, participants had a brief rest before proceeding to the formal experiment. They first completed three repetitions of the baseline scenario, followed by completing questionnaires about user experience and cognitive load in VR(as shown in Figure \ref{fig:teaser} left). Then, they performed tasks under two experimental conditions in a randomized order, with each condition repeated three times. After each set of tasks, participants completed the same questionnaires in VR. 
The process for each trial is shown in Figure \ref{fig:CognitiveLoadExperience}: participants pressed the left trigger on the controller to start the trial, then experienced a 3-second black screen with a countdown before the screen lit up, entering the corresponding experimental scenario. Participants completed the timed task, and groups with a cognitive task also performed the additional cognitive task. Participants pressed the left trigger to end the trial when they felt the target time had been reached.
Upon completing all experimental tasks, participants removed the equipment. They joined in a brief semi-structured interview, covering topics such as motion sickness symptoms, their subjective task performance, and their subjective experience of the scenarios.

\subsubsection{Data Collection}
There are three parts of data to be collected:

\textbf{Time Perception.}
We collected the following three types of time perception data to comprehensively assess participants' time perception under different experimental conditions: 
\begin{itemize}
    \item \textbf{Subjective time estimation ($t$):} This refers to the time that participants subjectively perceived during the 60s timing task. It is calculated by subtracting the start time of the task (after the countdown) from the time when the participant presses the left trigger to end the trial.
    \item \textbf{Average Subjective Time Estimation ($\hat{t}$):} This represents the participant's average perceived time across three repeated trials under the same experimental condition for the 60s timing task. It provides an overall measure of the participant's time perception under specific experimental conditions.
    \item \textbf{Relative Time Perception ($\Delta\hat{t}$): }This represents the change in time perception relative to the baseline under experimental conditions. It is calculated by subtracting the average subjective time estimation of the baseline condition from that of the experimental condition. The formula is as follows:
    \begin{equation}
     \Delta\hat{t}=\hat{t}_{Condition} - \hat{t}_{Baseline} 
    \end{equation}  
\end{itemize}

We define underestimation of time as when participants' subjective time perception is shorter than actual time, meaning participants spent more than 60 seconds in the scene ($t > 60 \ \text{or} \ \hat{t} > 60$). Overestimation of time occurs when subjective time is longer than actual time. Relative underestimation of time is when more time is spent in the experimental scenario than in the baseline ($\Delta\hat{t}>0$), while relative overestimation of time is the opposite.

\textbf{User Experience.}
After each experimental scenario, participants were asked to complete a questionnaire in VR. The user experience part of the questionnaire was adapted from the work of Katy Tcha-Tokey et al.~\cite{tcha2016questionnaire}, tailored to fit the specific design of our experiment. The questionnaire includes the following dimensions: presence ($PQ$), engagement ($UES$), immersion ($ITQ$), and emotion ($EQ$), with the emotion dimension containing one question and the other dimensions comprising three questions each. The total score of these dimensions represents the overall user experience score ($UEQ$). And we recorded cognitive load ($TLX$) with 2 questions~\cite{krieglstein2023development}. The questionnaire utilized a 7-point Likert scale~\cite{louangrath2018validity} for scoring. We also calculated relative user experience and cognitive load by subtracting the baseline condition scores from those obtained under the experimental scenario for each dimension.

\textbf{Physiological Data.}
We collected physiological data for each trial by the EEG-based multimodal physiological data device for modeling participants’ time perception. 
The recorded data includes raw EEG ($EEG$) and HR data ($HR$). Additionally, using the SDK provided by the equipment, we further recorded participants' brainwave data ($BW$, including $\alpha$, $low\beta$, $high\beta$, $\theta$ and $\gamma $), and human state data ($HS$), including stress, awareness, drowsiness, and meditation~\cite{han2022patent,han2021patent}. We also recorded the zeitgebers flag ($Zei$) corresponding to each physiological data.
\begin{figure*}[!h]
  \centering
  \includegraphics[width=0.9\textwidth]{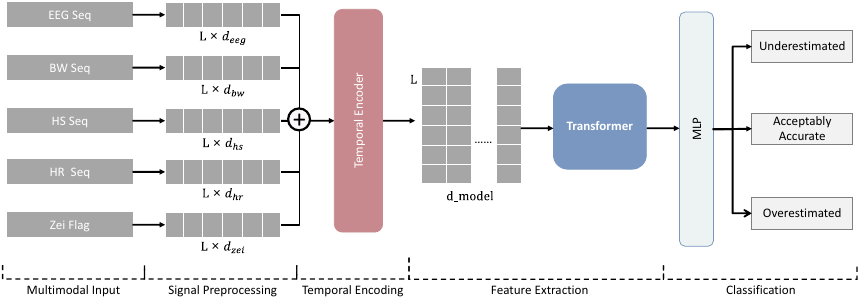}
  \caption{Overview of the data-driven TPM-Net architecture. The model consists of five stages. In the multimodal input stage, the BW Seq includes $\alpha$, low$\beta$, high$\beta$, $\theta$, and $\gamma$, while the HS Seq includes stress, awareness, drowsiness, and meditation. The Zei Seq represents the corresponding experimental variables. $d_{eeg}$, $d_{bw}$, $d_{hs}$, $d_{hr}$, $d_{zei}$ denotes the data channels corresponding to each sequence. Details of the temporal encoding module can be found in Figure \ref{fig:convolution}. }
  \label{fig:structure}
\end{figure*}
\begin{figure}
  \centering
  \includegraphics[width=0.45\textwidth]{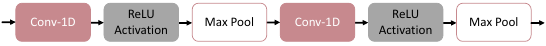}
  \caption{Temporal encoding module.}
  \label{fig:convolution}
\end{figure}
\subsection{Design of TPM-Net Architecture}
To better model time perception in VR, we propose the TPM-Net (as shown in Figure \ref{fig:structure}), which combines CNN and Transformer architectures tailored to the characteristics of physiological data. The collected physiological data is a set of 1D time-series sequences with different dimensions containing rich local temporal features and global dependencies. CNN is employed for fine-grained feature extraction and transient signal changes in multimodal signals ~\cite{yildirim2020deep}, while the Transformer captures long-range dependencies across sequences, modeling global relationships in the data~\cite{behinaein2021transformer,yang2022mobile}. By integrating CNN and Transformer, TPM-Net effectively captures both local and global features, improving predictive accuracy for complex multimodal temporal data. As shown in Figure \ref{fig:structure}, the TPM-Net consists of five stages: data input, signal prepossess, feature extraction, transformer processing, and a classification stage.

\textbf{Multimodal input.} In this study, we focus on four different types of 
physiological sequence data: raw EEG data, brainwave data(including $\alpha$, $low\beta$, $high\beta$, $\theta$, and $\gamma $), human state data (including stress, awareness, drowsiness, and meditation), and HR data. Additionally, zeitgeber flags were embedded into 4-dimensional vectors and concatenated with physiological data to enhance the model's capacity to capture factors influencing time perception. The length of each trial varies depending on the duration taken by participants.

\textbf{Signal preprocessing.} We first applied low-pass filtering and empirical mode decomposition to the raw EEG data to remove noise and baseline drift~\cite{zhang2008eeg}. Due to differences in sampling rates, we resampled the human state data using cubic spline interpolation to match the highest frequency~\cite{zitouni2021arousal}. To eliminate potential individual differences caused by age, gender, and personality, we normalized all data~\cite{zitouni2021arousal}. Finally, we padded the sequences within each batch to ensure consistent sequence lengths before feeding into TPM-Net. 

\textbf{Temporal encoding.} 
To extract more expressive temporal features from physiological and zeitgebers data, we applied the CNNs structure for each signal channel, capturing fine-grained temporal patterns through locally perceptive convolutional kernels and weight-sharing translation equivariance~\cite{chetana2023application}. As illustrated in Figure~\ref{fig:convolution}, the temporal encoding module consists of two parts, each starts with a 1D convolutional layer ($1\times3$ padded convolution), followed by a rectified linear unit (ReLU) and a Max Pooling layer. 

\textbf{Feature extraction.} 
We first fuse the features of multiple signal channels and represent them as a unified model embedding to capture latent time perception information, specifically formulated as:
\begin{equation}
    E=[E_{eeg},E_{bw},E_{hs},E_{hr},E_{zei}],
\end{equation}
where $E_{eeg}$, $E_{bw}$, $E_{hs}$, $E_{hr}$, and $E_{zei}$ represent the feature embeddings of EEG, BW, HS, HR, and Zei, respectively. This results in an $L \times d\_model$ ($d\_model = 128$) channel feature matrix, which is then fed into the Transformer module. The Transformer functions as the principal feature extractor, efficiently modeling both local relationships and long-range dependencies within and across the multimodal signals..

\textbf{Classification stage.} The features extracted by the Transformer are passed through a multilayer perception(MLP), yielding the final classification results. We categorized the labels into three groups: underestimated ($t \ge 69$), acceptably accurate ($51 \le t < 69$), and overestimated ($t < 51$). Physical time was chosen as the label because using relative time would lead the model to learn scenario-specific associations tied to the corresponding baseline, limiting its generalization and ability to capture the broader relationship between multimodal input and time perception beyond the experimental context. 
However, accurately estimating physical time is inherently challenging~\cite{walker2022impact}. In light of this, and given the standard deviation of the post-experimental data (STD = 10.25), which reflects the variability in participants' time estimations, we selected a $\pm 15\%$ empirical error margin~\cite{nesselroade2013handbook}. This error margin corresponds to the typical variation observed in participants' responses and allows for a reasonable tolerance in estimating physical time.


\section{Experimental Results}
\begin{figure*}[h]
  \centering
  \includegraphics[width=\textwidth]{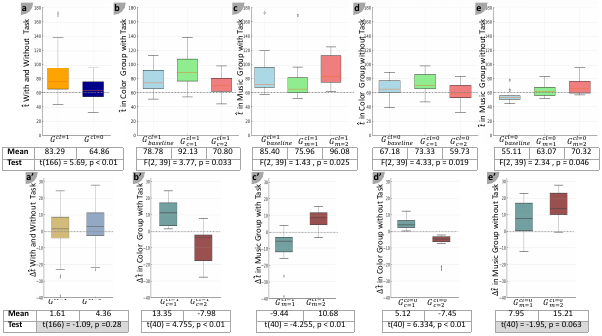}
  \caption{Subjective time perception results with means and data test. (a) results of subjective time estimation with/without task. (b) results of subjective time estimation with task in color group. (c)results of subjective time estimation with task in music group. (d) results of subjective time estimation without task in color group. (d)results of subjective time estimation without task in music group. (a'), (b'), (c'), and (d') are the results of corresponding relative time estimates to baseline as described above.   }
  \label{fig:PicOfAllData}
\end{figure*}
\subsection{Results for RQ1}
The Shapiro-Wilk test indicated that the $\hat{t}$ and the $\Delta \hat{t}$ for $G_{}^{cl=1}$ did not follow a normal distribution. As $\hat{t}$ and $\Delta \hat{t}$ for $G_{}^{cl=1}$ and $G_{}^{cl=0}$ come from different participants, we employed an independent samples t-test to compare these groups. For the remaining groups, we assessed the data distribution and confirmed normality. For $\hat{t}$, where each condition's data consists of three measurements from the same participant, we conducted Repeated Measures ANOVA for statistical testing. For $\Delta \hat{t}$, with each condition having two measurements from the same participant, we used paired samples t-test. Figure~\ref{fig:PicOfAllData} presents the mean values and test results.

\textbf{Task factor.} 
As shown in Figure~\ref{fig:PicOfAllData} part a and a', when the effects of color and music tempo are ignored, $\hat{t}$ is significantly more significant in the task group than in the non-task group ($p<0.01$). However, there is no significant difference in $\Delta \hat{t}$ between the two groups ($p>0.05$). Regardless of the presence of tasks, 71.35\% of participants tended to underestimate time, with this tendency being more pronounced in the task group, where 86.42\% of participants underestimated time.

\textbf{Visual factor(light color).}
As shown in Figure~\ref{fig:PicOfAllData} part b, b', d and d', in the task group, participants under all lighting colors tended to underestimate time ($\left | \overline{\hat{t}}  \right | _{c=0/1/2}^{cl=1}>60 $), with red lighting leading to the most significant underestimation ($\left | \overline{\hat{t}}  \right | _{c=1}^{cl=1} > \left | \overline{\hat{t}}  \right | _{c=0}^{cl=1}>\left | \overline{\hat{t}}  \right | _{c=2}^{cl=1}$). In terms of relative time estimation, participants in the red lighting scenario exhibited a tendency toward relative underestimation ($\left | \overline{\Delta \hat{t}}  \right | _{c=1}^{cl=1}>0 $), while those in blue lighting tended to relatively overestimate time ($\left | \overline{\Delta \hat{t}}  \right | _{c=2}^{cl=1}<0 $). In the non-task group, participants under blue lighting tended to overestimate time ($\left | \overline{\hat{t}}  \right | _{c=2}^{cl=0}<60 $), while under other colors, they tended to underestimate time. The relative time estimation results were consistent with those in the task group.

\textbf{Auditory factor(music tempo).}
As shown in Figure~\ref{fig:PicOfAllData} part c, c', e and e',  in the task group, participants under all music tempo conditions generally underestimated time ($\left | \overline{\hat{t}}  \right | _{m=0/1/2}^{cl=1}>60 $), with the most pronounced effect observed under slow tempo music ($\left | \overline{\hat{t}}  \right | _{m=2}^{cl=1} > \left | \overline{\hat{t}}  \right | _{m=0}^{cl=1}>\left | \overline{\hat{t}}  \right | _{m=1}^{cl=1}$).   In contrast, participants were more likely to overestimate time under fast-tempo music than under the no-music condition. 
In the non-task group, the absence of music led to a greater tendency to overestimate time ($\left | \overline{\hat{t}}  \right | _{m=0}^{cl=1} < 60$). In contrast, slow tempo music tended to result in underestimation ($\left | \overline{\hat{t}}  \right | _{m=2}^{cl=1} >\left | \overline{\hat{t}}  \right | _{m=1}^{cl=1}> 60$).
In the task group, fast tempo music led to relative overestimation of time ($\left | \overline{\Delta \hat{t}}  \right | _{m=1}^{cl=1}<0 $), while slow tempo music resulted in relative underestimation ($\left | \overline{\Delta \hat{t}}  \right | _{m=2}^{cl=1}>0 $); in the non-task group, there was no significant difference in relative time estimation ($p>0.05$).

\subsection{Results for RQ2}
\begin{figure}[!b]
     \centering
  \begin{center}
\includegraphics[width= 0.8\columnwidth]{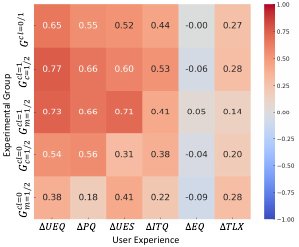}
  \caption{Heatmap of correlation between relative subjective time perception and relative user experience.}
  \label{fig:Correlation}
    \end{center} 
\end{figure}

\begin{table*}[h]
  \centering
  \caption{Correlation between $\hat{t}$ and User Experience and its subcomponents, and correlation between $\Delta \hat{t}$  and Relative User Experience and its subcomponents. We tested the correlations using Pearson's correlation coefficient, where $r$ denotes the correlation coefficient, $p$ indicates the significance level, and $R_{}^{2}$ represents explanatory power.}
    \begin{tabular}{p{6.5em}ccc|ccc}
    \toprule
    \multicolumn{1}{c}{\multirow{2}[2]{*}{}} & \multicolumn{3}{c}{\textbf{Correlation between $\hat{t}$ and User Experience}} & \multicolumn{3}{c}{\textbf{Correlation between $\Delta \hat{t}$ and Relative User Experience}} \\
    \multicolumn{1}{c}{} & \multicolumn{1}{p{6.375em}}{\makecell[c]{r} } & \multicolumn{1}{p{6.375em}}{\makecell[c]{p}} & \multicolumn{1}{p{6.375em}}{\makecell[c]{R²}} & \multicolumn{1}{p{6.375em}}{\makecell[c]{r}} & \multicolumn{1}{p{6.375em}}{\makecell[c]{p}} & \multicolumn{1}{p{6.375em}}{\makecell[c]{R²}} \\
\cmidrule{2-7}    UX    & 0.437 & 2.31E-09 & 0.191 & 0.651 & 4.49E-15 & 0.424 \\
    Presence & 0.408 & 2.97E-08 & 0.167 & 0.554 & 1.56E-10 & 0.307 \\
    Engagement & 0.446 & 9.51E-10 & 0.199 & 0.524 & 2.18E-09 & 0.275 \\
    Immersion & 0.385 & 2.05E-07 & 0.148 & 0.442 & 8.27E-07 & 0.196 \\
    Emotion & 0.26  & 5.93E-04 & 0.068 & -0.003 & 9.76E-01 & 0.000  \\
    Cognitive Load & 0.315 & 5.13E-05 & 0.099 & 0.302 & 4.32E-04 & 0.091 \\
    \bottomrule
    \end{tabular}%
  \label{table:corelationR2}%
\end{table*}%
We first calculated the correlation between $\hat{t}$ and $UEQ$ and its subcomponents using Pearson's correlation coefficient and tested for significance. We then applied the same method to analyze the correlation between $\Delta \hat{t}$ and relative user experience, including its subcomponents. Additionally, we calculated the coefficient of determination ($R_{}^{2}$) of both methods (as shown in Table~\ref{table:corelationR2}). Finally, we conducted a detailed analysis of the relationship between $\Delta \hat{t}$, $\Delta UEQ$, subcomponents of user experience, and $TLX$ (as shown in Figure~\ref{fig:Correlation}).

\textbf{Research methods.} According to the data in Table~\ref{table:corelationR2}, the correlations between $\Delta \hat{t}$ and $\Delta UEQ$ (r = 0.651), including subcomponents of user experience, are significantly more potent than those between $\hat{t}$ and $UEQ$ (r = 0.437). Additionally, the RSTC method demonstrates higher explanatory power. Specifically, the RSTC method shows greater significance (smaller p-values) and more substantial explanatory power (higher $R_{}^{2}$ values) for $UEQ$, $PQ$, and $UES$. Although the p-value is more significant for $ITQ$, the $R_{}^{2}$ value is still higher in the RSTC method. However, the RSTC method results in more substantial p-values and smaller $R_{}^{2}$ values for emotion and cognitive load.

\textbf{Relative time perception and relative user experience. } Overall, $\Delta \hat{t}$ showed a significant positive correlation with $\Delta UEQ$ ($r = 0.65, p < 0.001$) and also with $\Delta PQ$ ($r_{\Delta PQ} = 0.55$) and $\Delta UES$ ($r_{\Delta UES} = 0.52$). The correlation with $\Delta ITQ$ ($r_{\Delta ITQ} = 0.44$) and $\Delta TLX$ ($r_{\Delta TLX} = 0.27$) was weaker, and there was almost no correlation with $\Delta EQ$ ($r_{\Delta EQ} = -0.00$). 
Across different experimental groups, $\Delta \hat{t}$ in $G_{c=1/2}^{cl=1}$ showed significant positive correlations with $\Delta UEQ$, $\Delta PQ$, $\Delta UES$, and $\Delta ITQ$. In $G_{m=1/2}^{cl=1}$, $\Delta \hat{t}$ was positively correlated with $\Delta UEQ$, $\Delta PQ$, and $\Delta UES$. In $G_{c=1/2}^{cl=0}$, it correlated positively with $\Delta UES$ and $\Delta PQ$. However, in $G_{m=1/2}^{cl=0}$, there were no significant correlations. Across all groups, $\Delta \hat{t}$ showed no significant correlation with $\Delta EQ$ or $\Delta TLX$.

\subsection{Results for RQ3}
\textbf{Physiological dataset. }After removing invalid data due to collection errors or user input issues (e.g., noise or incomplete data), 477 valid instances were retained. The final dataset included 243 underestimation, 194 acceptable accuracy, and 49 overestimation instances.

\textbf{Implementation details.} 
We implemented our proposed approach using the PyTorch framework, with a learning rate set to 5e-4 and a batch size of 24. The Adam optimizer with weight decay ($\lambda $ = 5e-4) was employed to update the model's parameters. Additionally, dropout was utilized to mitigate the risk of overfitting. We used an 85\%-15\% split for training and testing to ensure the model had sufficient data to learn complex patterns while the test set remained adequate to evaluate generalization. Furthermore, we allocated 20\% of the training data as a validation set to fine-tune the model during training. The model was trained for 300 epochs, with early stopping applied (patience = 40) to prevent overfitting. Additionally, we employed 5-fold cross-validation to ensure a more robust evaluation of the model's performance.

\textbf{Model performance and comparison results.} 
To validate the feasibility of modeling time perception based on physiological data in VR environments, we implemented the TPM-Net architecture described in section 3.3. As shown in Table \ref{tab:zhibiao}, the TPM-Net achieved high performance in predicting time perception from multimodal physiological data (Accuracy = 86.11\%, Macro-F1 = 83.10\%, UAR = 79.45\%). To evaluate the performance of TPM-Net, we compared it with five commonly used baseline models: (1) Naive Bayes (NB)~\cite{wang2016detection}, (2) Support Vector Machine with linear kernel (SVM-LR)~\cite{subramanian2016ascertain}, (3) Support Vector Machine with radial basis function kernel (SVM-RBF)~\cite{subramanian2016ascertain}, (4) CNN~\cite{yildirim2020deep}, and (5) bidirectional Long Short-Term Memory network (BiLSTM)~\cite{zitouni2021arousal}. All models were retrained on the same training-testing set splits to ensure fair comparison, and the results (Table \ref{tab:zhibiao}) demonstrate that the TPM-Net outperforms the other methods across all metrics, affirming its effectiveness in predicting time perception in VR environments. 
\begin{table}[!h]
  \centering
  \caption{Comparison of different approaches. Macro-F1 represents the average F1 score, and UAR represents the unweighted average recall~\cite{Quan2021}.}
  \scalebox{0.95}{
    \begin{tabular}{lccc}
    \toprule
    \textbf{Approach} & \textbf{Accuracy(\%)} & \textbf{Macro-F1(\%)} & \textbf{UAR(\%)} \\
    \midrule
    NB    & 45.83    & 36.60    & 44.88 \\
    SVM-LR   & 38.89    & 37.34    & 41.69 \\
    SVM-RBF   & 54.17    & 44.87    & 43.71 \\
    CNN   & 65.62    & 46.22    & 49.05 \\
    BiLSTM  & 65.28    & 45.93    & 50.48 \\
    \textbf{TPM-Net} & \textbf{83.33}    & \textbf{74.84}   & \textbf{71.32} \\
   
    \bottomrule
    \end{tabular}%
    }
  \label{tab:zhibiao}%
\end{table}%

\subsection{Results for Semi-structured Interview}
One participant experienced significant cybersickness after the experiment, leading to the exclusion of his data. The remaining participants reported no strong negative emotions or discomfort in the color scenarios. All participants found the experiment's music comfortable without any strong feelings of preference or dislike.

\section{Discussion}
\subsection{Discussion about RQ1}
 Most participants tended to underestimate time, especially in the group with tasks. 
 This underestimation occurred because, in the dual-task model~\cite{navon1979economy}, participants had to share attentional resources between the timing and spatial rotation tasks. This division of attention led to mutual interference, causing time underestimation. Conversely, in waiting scenarios, participants felt time passed more slowly, which is also consistent with attention resource theory~\cite{ barrouillet2007time}.
 The novelty of VR, particularly for participants with less experience, have amplified this underestimation~\cite{huang2020investigating}. Furthermore, the lack of significant differences in relative time estimation between task and non-task groups suggests that the RSTC method effectively bridges the gap between different experimental factors and physiological variables such as age and gender.
For visual zeitgeber (lighting color), the results indicate that red lighting is more likely to lead to time underestimation, while blue lighting causes time overestimation in the groups with tasks. Compared to the baseline scenario (white light), red light leads to relative time underestimation, while blue light leads to relative time overestimation. This phenomenon can be explained by the sensitivity of intrinsically photosensitive retinal ganglion cells (ipRGCs) to blue light, which increases subjective arousal and attention, thereby expanding subjective duration~\cite{yang2018subjective}. This also demonstrates that the simulation of colored light in VR aligns with its effects on time perception in real-world environments.

For auditory zeitgeber (music tempo), the presence of music had a significant impact on time perception in the groups without task. Regardless of tempo, participants tended to underestimate the time when music was present, likely due to music diverting attention in the absence of tasks~\cite{rogers2020potential}. However, in the group with the task, slow-tempo music led to more significant time underestimation, while the opposite was true in groups without the task. This might be due to the entrainment effect~\cite{ross2022time} being more prominent without tasks. Additionally, when combined with fast-tempo music, the sandbag's slower animation and refresh rate in the task might have caused a mismatch in rhythm, further distracting participants~\cite{phillips2014composer,landeck2023clocks}.
\subsection{Discussion about RQ2}
Previous research often compares participants' time estimates to physical time~\cite{liao2020data, landeck2023clocks,schatzschneider2016turned}, which presents particular challenges when studying the relationship between time perception and user experience. On the one hand, few individuals can accurately estimate physical time~\cite{walker2022impact}, and on the other hand, physical time does not necessarily equate to a standard user experience. To address these challenges, we validated two approaches: (1) directly analyzing the relationship between participants' subjective time estimates and user experience and (2) using the RSTC method, which compares participants' time perception in different experimental scenarios relative to a baseline scenario and analyzes its correlation with user experience. The results show that the RSTC method provides more substantial significance and higher explanatory power ($R_{}^{2}$ value) in validating the significant positive correlations between $\Delta \hat{t}$ and relative overall user experience, presence, and engagement. This suggests a promising approach to infer overall user experience from relative time perception changes.

Applying the RSTC method, we found a positive correlation between participants' relative time perception changes and relative overall user experience—specifically, the more participants underestimated time and stayed longer in the experimental environment, the better their experience. This finding aligns with everyday life~\cite{freedman2014does} and other mediums~\cite{ sackett2010you}. Relative time perception strongly correlated with overall user experience, presence, and engagement but showed weaker correlations with immersion, emotion, and cognitive load. Task settings and changes in visual/auditory stimuli significantly impacted presence and engagement.
Regarding emotion, although literature suggests that emotion influences time perception~\cite{gable2022does,nieuwoudt2014time}, our study did not find significant correlations. This may be due to the zeitgebers we selected, which may not have been sufficient to induce substantial emotional changes (81.74\% of participants chose ``neutral (4/7 on the 7-point scale)" or ``somewhat positive (5–6/7 on the 7-point scale)"). The impact of the cognitive load was also minimal, different from other study~\cite{nieuwoudt2014time}, likely because the task difficulty was not varied (85\% of participants rated the difficulty as moderate (7-8/14 on the 14-point scale)). 

The correlation between relative time perception changes and engagement was more pronounced in groups with the task. This suggests that designers can consider adjusting user engagement in task-oriented environments to influence time perception. In groups without the task, the correlation between relative time changes and relative user experience was weaker, particularly in the auditory condition. We believe that both task content and lighting changes provided participants with a wealth of visual information. In VR environments, visual stimuli typically impact users more than auditory stimuli~\cite{voinescu2020exploring}. This implies that effectively manipulating time perception to enhance user experience requires not just altering music content but also integrating visual information (including visual and task-related zeitgebers).

\subsection{Discussion about RQ3}
The results have shown that multimodal data-driven approaches can effectively model users' time perception in VR environments. In this study, we proposed the TPM-Net to model time perception based on multimodal physiological data, achieving promising accuracy. This demonstrates that multimodal physiological signals (e.g., EEG, HR, etc.) provide rich information for modeling time perception, and the combination of CNNs and the Transformer module effectively captures both local and global features of the multimodal data~\cite{yang2022mobile}. Compared to traditional models, the framework used in this study demonstrated superior performance, proving the feasibility of predicting time perception in VR environments. 
The classification labels are based on physical time rather than relatively subjective temporal changes, which gives this model good generalization ability. The differences between labels under different experimental conditions allowed us to infer relative time perception changes and, consequently, shifts in user experience. 
Due to the dataset's limited size, the labels' granularity was broad, restricting precise predictions of time perception. But it can provide an accurate trend judgment. 
Overall, deep learning methods based on multimodal physiological data show great potential for predicting time perception in VR environments. As dataset size and label precision improve, these methods will provide even higher accuracy.

\section{Conclusions and Future Work}
In this study, we explored how different zeitgebers—light color, music tempo, and task factors—affect time perception in VR environments. Through our proposed Relative Subjective Time Change (RSTC) method, we analyzed the relationship between relative changes in time perception and user experience, revealing that relatively underestimating time in VR is strongly associated with an improved overall user experience, presence, and engagement. Moreover, this research highlights the potential of multimodal physiological data combined with a deep learning method to model users' time perception accurately. The TPM-Net framework proposed in this study outperformed traditional methods in predicting time perception, paving the way for its applications in VR design, VR-based therapy, and specialized sector training. 

While this study provides valuable insights into time perception in VR, several areas can be further explored and improved in future work.
Firstly, the results revealed that when other zeitgebers are linked to tasks (e.g., matching music tempo with tasks), they have a stronger impact on time perception than non-task-linked environmental zeitgebers like light color. This suggests that future studies could further explore how task-linked zeitgebers influence time perception in VR and the mechanisms behind these effects, such as the color of task objects~\cite{gorn2004waiting} or task-linked animation speed~\cite{matthews2011changes}.
Secondly, the physiological data collected in this study could benefit from a larger dataset and more granular label classification, allowing for finer predictions and more precise inference of relative changes in time perception. 
Finally, while the TPM-Net framework shows promise, further exploration of advanced deep learning architectures~\cite{cheng2024novel,dutta2024emocomicnet} could improve the accuracy of time perception predictions and offer deeper insights into time perception in VR.

\acknowledgments{
The authors wish to thank A, B, and C. This work was supported in part by
a grant from XYZ.}

\bibliographystyle{abbrv-doi}
\normalem
\bibliography{Reference}
\end{document}